\documentclass[a4paper,12pt]{article}
\usepackage{amssymb,amsmath}
\def\beq{\begin{equation}}
\def\eeq{\end{equation}}
\def\bea{\begin{eqnarray}}
\def\eea{\end{eqnarray}}
\def\nn{\nonumber}
\def\U{{\cal U}}

\def\qb#1{\bigg[ #1 \bigg]}

\def\s#1{\bigg\lgroup #1 \bigg\rgroup}

\makeatletter
  \def\@cite#1#2{${\mbox{#1\if@tempswa , #2\fi}}$}
\makeatother

\makeatletter
  \def\@biblabel#1{$^{\mbox{#1}}$}
\makeatother
\begin{document}
%
%
%
%
\thispagestyle{empty}

\vspace*{3cm}
\begin{center}
{\LARGE\sf
Rigid rotators and diatomic molecules via Tsallis statistics}\\

\bigskip\bigskip
R. Chakrabarti, R. Chandrashekar and S.S. Naina Mohammed \\

\bigskip
\textit{
Department of Theoretical Physics, \\
University of Madras, \\
Guindy Campus, Chennai 600 025, India
}

\end{center}

\vfill
\begin{abstract}
We obtain an analytic expression for the  specific heat of a system of 
$N$ rigid rotators exactly  in the high temperature limit, and via a 
pertubative approach in the low temperature limit.  We then evaluate the 
specific heat of a diatomic gas  with both translational and rotational 
degrees of freedom, and conclude that there is a mixing between the 
translational and rotational degrees of freedom in nonextensive 
statistics. 
\end{abstract}
PACS Number(s): 05.20.-y, 05.70 
\setcounter{page}{1}
%
%
%
\setcounter{equation}{0}
\section{Introduction}
\label{Intro}
The nonextensive generalization of Boltzmann-Gibbs statistics was proposed by 
Tsallis [\cite{CT1}] to study systems involving long-range interactions,
long-range microscopic memory, and nonequilibrium phenomenon. The 
nonextensive statistics has been applied in a wide range of areas like 
self-gravitating systems [\cite{PP}, \cite{FL}, \cite{DJ}], solar 
neutrinos [\cite{KLQ1}, \cite{KLQ2}], and  biological systems 
[\cite{TBM}, \cite{AH}]. The functional form of the generalized entropy 
reads 
\beq
S_{q}= k \qb{\frac{W^{1-q}-1}{1-q}} \equiv k \ln_{q} W,  \qquad    q \in 
{\mathbb{R}_{+}},
\label{Tqe}
\eeq 
where $k$ is a constant generalizing the Boltzmann constant $k_{B}$ for 
an arbitrary $q$, and $W$ is the weight factor. In the $q \rightarrow 
1$ limit the Tsallis entropy $S_{q}$ reduces to the standard 
Boltzmann-Gibbs entropy $S$. It is interesting to observe that these 
two entropies may be interrelated via hypergeometric functions as
\bea
 S/k_{B} &=& {}_{2}F_{1} (1,1;2;-(1-q)S_{q}/k) \ S_{q}/k, \\
 \label{er1}
 S_{q}/k &=& {}_{1}F_{1} (1;2;(1-q)S /k_{B}) \ S/k_{B}.
 \label{er2}
\eea
Based on the entropy (\ref{Tqe}) a canonical formulation was developed
in [\cite{CUR}, \cite{TMP}]. As a concrete example of using this 
formalism for studies of various statistical systems at equilibrium the 
classical monoatomic ideal gas has been investigated as a $N$-body 
problem [\cite{SA1}, \cite{AMPP}] in detail, and, in particular, the 
specific heat of this system has been computed exactly. 
\par 

In the above context study of the rigid rotator has been initiated in  
[\cite{BCCT}, \cite{BST}]. Using molecular dynamics a numerical study of 
$N$ rotators with a long-range interaction has been done in 
[\cite{BCCT}], where a long-standing metastable state with anomalous 
duration suggesting nonextensivity has been observed. In [\cite{BST}] an 
analytical study of the rigid and non-rigid rotators has been done. 
These authors, however, obtained the specific heat of the system by 
studying a single rigid rotator. In the regime of nonextensive 
statistics, strictly speaking, the study of a single rigid rotator cannot 
be extended to a system of $N$ rigid rotators as it is familiarly done 
in the extensive Boltzmann-Gibbs scenario. A systematic study of 
statistics of $N$ rigid rotators is, therefore, essential for 
understanding the dependence of the specific heat on the temperature. 
We will comment on the other relevant issues in appropriate places.

\par

In the present work we first evaluate the specific heat at the high 
temperature limit for $N$ rigid rotators, and a diatomic gas endowed 
with both translational and rotational degrees of freedom. Towards this 
end we construct an implicit equation involving the generalized partition 
function and the internal energy. In the high temperature limit the 
implicit equation is solved \textit {exactly} and explicit expression 
for the specific heat is evaluated. Contrarily, obtaining an exact 
solution of the said implicit equation for a system of $N$ rigid 
rotators in the low temperature limit appears to be difficult.
We, therefore, use the method developed in [\cite{TMP}] that 
interrelates the ensemble probabilities obtained via the second 
and the third constraint approaches, respectively, after performing a 
suitable transformation on the temperature. This transformation connects 
the second constraint approach [\cite{CUR}] that utilizes unnormalized 
$q$-expectation values to the third constraint [\cite{TMP}] approach 
based on the normalized $q$-expectation values. To construct the 
pertinent transformation here we adopt a pertubative approach by 
disentangling the $q$-exponential in an infinite series of ordinary 
exponential [\cite{CQ}], and retain terms upto the order $(1 - q)^{2}$. 
In the context of the gas of $N$ diatomic molecules we notice that a 
mixing of the translational and rotational degrees of freedom in the 
nonextensive terms occur for the specific heat. We here assume a large 
mass for the diatomic molecules so that their vibratory motions 
may be neglected. The plan of the paper is as follows. The specific heat 
of $N$ rigid rotators in the high and low temperature regimes are 
discussed in Sections \ref{HTL} and \ref{LTL}, respectively. This is 
followed by our discussions on the specific heat of a diatomic gas in 
Sec. \ref{DM}. Our concluding remarks are given in Sec. \ref{CR}.

%
%
%
\setcounter{equation}{0}
\section{ Rigid rotators in the high temperature limit}
\label{HTL}
The energy eigenvalues of an ensemble of $N$ isotropic rigid rotators 
read 
\beq
E = \frac{\hbar^{2}}{2I} \sum_{i = 1}^{N} l_{i}(l_{i}+1), \qquad 
l_{i} = 0, 1, \hdots,
\label{eer}
\eeq
where $I$ is the moment of inertia of a rotator. The generalized 
partition function of a canonical ensemble of $N$ rotators at an
inverse temperature $\beta=1/k T$ in the third constraint framework
[\cite{TMP}] is given by 
\beq
\bar{Z}^{(3)}_{q} (\beta) = \sum_{l_{i}=0}^{\infty} {\cal{D}} 
               \qb{1-(1-q) \frac{{\hat{\beta}}}{c}
                ({\cal{L}}- \Omega_{R} U_{q})}^{\frac{1}{1-q}}, \qquad
c=\sum_{i}[p_{i}^{(3)}(\beta)]^{q},
\label{gpfr}
\eeq
where $\Omega_{R}=2I/\hbar^{2}, {\hat{\beta}}=\beta/\Omega_{R}$. The
degeneracy factor ${\cal{D}}$ and the sum ${\cal L}$ read
\bea
{\cal{D}} = \prod_{i=1}^{N} (2l_{i}+1),  \qquad
{\cal{L}} =  \sum_{i=1}^{N} l_{i}(l_{i}+1).
\label{dl}
\eea 
In the expression (\ref{gpfr}) the ensemble probability in the third 
constraint approach is given by
\beq
p_{i}^{(3)}(\beta) =  \qb{1-(1-q) \frac{{\hat{\beta}}}{c}
                    ({\cal{L}}- \Omega_{R} U_{q})}^{\frac{1}{1-q}} 
                    \;\;(\bar{Z}^{(3)}_{q} (\beta))^{-1}. 
\label{P3}
\eeq
At high temperature limit the rotational levels are closely spaced,  
and, therefore, the sum in (\ref{gpfr}) is well-approximated by an 
integral
\beq
[\bar{Z}^{(3)}_{q} (\beta)] =  \int_{l_{i}=0}^{\infty} {\cal{D}} 
                               \qb{1-(1-q) \frac{{\hat{\beta}}}{c}
                               ({\cal{L}}- \Omega_{R} U_{q})}^{\frac{1}{1-q}}
                               \prod_{i=1}^{N} {\rm{d}}l_{i}.
\label{gpfri}
\eeq
The integral may be exactly evaluated as
\beq
\bar{Z}^{(3)}_{q} (\beta) = c^{N} \ \qb{\exp_{q}\bigg( \frac{\beta}
                           {c} U_{q}^{(3)}\bigg)} ^{\Lambda_{(1)}}
                           \;\;\Phi\;\;{\hat{\beta}}^{-N}, \qquad  
                             \Phi = \prod_{n=1}^{N}[1+(1-q)n]^{-1},
\label{gpfrs}
\eeq
where $\Lambda_{(k)}=1+ k (1-q)N$. As a 
consequence of the property [\cite{TMP}] 
\beq
c=[\bar{Z}^{(3)}_{q} (\beta)]^{(1-q)}.
\label{CZ}
\eeq
the above expression of the generalized partition function may be 
regarded as an implicit equation. 
The internal energy of the system introduced via the escort 
probabilities [\cite{TMP}]
\beq
U_{q}^{(3)} = \frac{\sum_{i} E_{i} (p_{i}^{(3)})^{q}}
{\sum_{i} (p_{i}^{(3)})^{q}}
\label{ie3}
\eeq
may be recast in the following form: 
\beq
U_{q}^{(3)} = \frac{1}{\Omega_{R} \bar{Z}^{(3)}_{q} (\beta)}
              \int_{l_{i}=0}^{\infty} {\cal{L}} \ {\cal{D}} 
              \qb{1-(1-q) \frac{{\hat{\beta}}}{c}
              ({\cal{L}}- \Omega_{R} U_{q})}^{\frac{q}{1-q}}
              \prod_{i=1}^{N} {\rm{d}}l_{i}. 
\label{iei}
\eeq
Parallel to (\ref{gpfri}) the above integral may be completed, and 
this, in turn, produces an implicit equation for the internal energy:  
\beq
U_{q}^{(3)} = \frac{N c^{N+1}}{\beta \bar{Z}^{(3)}_{q} (\beta)} \ \Phi \
               \qb{\exp_{q}\bigg(\frac{\beta}{c} U_{q}^{(3)}\bigg)} 
              ^{\Lambda_{(1)}}{\hat{\beta}}^{-N}.
\label{ies}             
\eeq
Substituting (\ref{gpfrs}) in (\ref{ies}) the internal energy is found 
to be proportional to the trace $c$: 
\beq
 U_{q}^{(3)} = c \ N \ \beta^{-1}.
\label{ratio}
\eeq
Employing (\ref{gpfri}), (\ref{CZ}) and (\ref{ratio}) we obtain an 
explicit expression for the generalized partition function:
\beq
\bar{Z}^{(3)}_{q} (\beta) = [\exp_{q}(N)]^{\mu_{1}} 
                                \ \Phi^{\mathfrak{L}_{(1)}} 
\; \hat{\beta}^{-N \mathfrak{L}_{(1)}}, \qquad
\mathfrak{L}_{(k)} = {\Lambda_{(-k)}^{-1}}, \qquad
\mu_{k} = \Lambda_{(k)}\mathfrak{L}_{(k)} .
\label{gpfef}
\eeq
The structure of $\Phi$ given in (\ref{gpfrs}) makes it evident that 
the generalized partition function (\ref{gpfef}) for $N$ rotators  
has singularities at the values of $q$ given by $q=(n+1)/n$ 
for $n=1,2,...,N$. The number of singularities equals the number of 
degrees of freedom of the system. As the number of rotators $N$ 
increases the singularities tend to accumulate at $q=1$, the limiting 
value where the statistical mechanics becomes  extensive. The 
generalized partition function is, therefore, well-defined in the 
interval $0 < q < 1 + 1/N$, where the upper limit is characterized by 
the total number of the rotators. The above discussion makes it 
apparent that in the context of nonextensive thermodynamics the size 
$N$ of the system is of key importance in selecting the admissible 
range of the nonextensivity parameter $q$. In contrast to our result 
the authors of Ref. [\cite{BST}] considered the generalized partition 
function at the high temperature limit for the parameter range 
$1 < q < 2$. Employing the results (\ref{CZ}), (\ref{ratio}) and 
(\ref{gpfef}) the internal energy  
\beq
U_{q}^{(3)}(\beta) = \frac{N}{\beta} \ \Lambda_{(1)}^{\mu_{1}} \  
                     \Phi^{(1-q) \mathfrak{L}_{(1)}}
                      \ \hat{\beta}^{(q-1)N \mathfrak{L}_{(1)}}.             
\label{ieef}
\eeq
and the specific heat of the system may be computed:
\beq
C_{q}^{(3)} \equiv  \frac{\partial U_{q}^{(3)}}{\partial T}
            =   N \ k \  {\mathfrak{L}_{(1)}} \ \Phi^{(1-q) \mathfrak{L}_{(1)}}
               \  \Lambda_{(1)}^{\mu_{1}} \
              {\hat{\beta}}^{{(q-1)N \mathfrak{L}_{(1)}}}.
\label{shht}
\eeq
The specific heat of the rigid rotators in the high temperature limit, 
in contrast to its classical value, becomes a temperature dependent 
quantity for nonextensive statistics. In the extensive $q \rightarrow 1$ 
limit we recover the classical Boltzmann-Gibbs statistics for these 
thermodynamics quantities: 
$U_{q} \rightarrow N \ k_{B}T,\;\;C_{q} \rightarrow N \ k_{B}$. 

\par

To obtain the internal energy (\ref{ieef}) and the specific 
heat (\ref{shht}) we may also employ the method discussed in 
[\cite{TMP}], where the physical quantities evaluated in the second 
constraint approach are transformed to those in the third constraint 
approach by introducing a fictitious temperature $\beta^{\prime}$ that 
is useful for calculation. The ensemble probabilities in the second 
constraint and the third constraint approaches are now linked as
\beq
p^{(3)} (\beta) = p^{(2)} (\beta^{\prime}). 
\label{PEQ}
\eeq
In other words, the second constraint approach is employed as a 
convenient tool at the intermediate stages of calculation, and finally 
results are translated to the third constraint approach via certain 
transformations. The general transformation rule for the temperature reads 
\beq
\beta = \beta^{\prime} \frac{\sum_{i}[p_{i}^{(2)}(\beta^{\prime})]^{q}}
        {1-\frac{(1-q) \beta^{\prime} U_{q}^{(2)}(\beta^{\prime})}
        {\sum_{i}[p_{i}^{(2)}(\beta^{\prime})]^{q}}}
\label{trans}
\eeq
In the present example of the rigid rotators at the high 
temperature limit this relation assumes the form
\beq
\frac{1}{\beta^{\prime}} =  \qb{\Omega_{R}^{(1-q)} \  \Phi^{(1-q)} \
                            \Lambda_{(N)}^{2} }^{\mathfrak{L}_{(1)}} 
                            \s{\frac{1}{\beta}}^{\mathfrak{L}_{(1)}}.
\label{rtrans}
\eeq
The specific heat and the internal energy evaluated using this 
technique is identical to the results (\ref{ieef}) and (\ref{shht})
where the third constraint method has been used \textit {ab initio}.

%
%
\setcounter{equation}{0}
\section{ Rigid rotators in the low  temperature limit}
\label{LTL}
In the example of $N$ rigid rotators at the low temperature limit we 
obtain the thermodynamic quantities perturbatively by treating 
$(1 - q)$  as the series parameter. This process involves use of the 
second constraint approach as an intermediate step. The results 
corresponding to the third constraint approach are then extracted 
[\cite{TMP}] via transformations (\ref{PEQ}) and (\ref{trans}) 
linking the ensemble probabilities in these 
two approaches. The partition function of $N$ rigid rotator in the 
second constraint setting reads 
\beq
 Z_{q}^{(2)} (\beta)  \equiv Tr \exp_{q} (- \hat{\beta}\;{\cal L}) 
= \sum_{l_{i}=0}^{\infty} {\cal{D}} \
               [1-(1-q){\hat{\beta}} {\cal{L}} ]^{\frac{1}{1-q}},
\quad \exp_{q} ({\cal X}) = (1 + (1 - q) \;{\cal X})^{\frac{1}{1 - q}}
\label{Z2}
\eeq
and the pertinent ensemble probability is given by 
\beq
p^{(2)}(\beta) = [1-(1-q){\hat{\beta}} {\cal{L}} ]^{\frac{1}{1-q}}
\; (Z_{q}^{(2)} (\beta))^{- 1}.  
\label{P2DEF}
\eeq
In the low temperature regime the $l_{i}=0$ and $l_{i}=1$ quantum 
levels generate the dominant contribution to the partition function.

\par
 
Using an infinite product expansion of the  $q$-exponential [\cite{CQ}]  
\beq
\exp_{q}(-{\hat{\beta}}{\cal L}) = \exp\bigg( - \sum_{k=1}^{\infty} 
                           \frac{(1-q)^{k-1}}
                           {k} {\hat{\beta}}^{k} {\cal{L}}^{k} \bigg)
\label{dqe}
\eeq
we express the partition function (\ref{Z2}) as a pertubative series in 
$(1-q)$, and retain terms upto second order:
\beq
Z_{q}^{(2)}(\beta) = \sum_{l_{i}=0,1} {\cal{D}} \qb{1-(1-q) 
              \frac{{\hat{\beta}}^{2}}{2} {\cal{L}}^{2} + (1-q)^{2} 
              \bigg(\frac{{\hat{\beta}}^{4}}{8} {\cal{L}}^{4} -
              \frac{{\hat{\beta}}^{3}}{3} {\cal{L}}^{3} \bigg)}
              \exp(-{\hat{\beta} {\cal{L}}}).
\label{rpps}
\eeq
Recasting the above series as 
\beq
Z_{q}^{(2)}(\beta) = \qb{ 1-(1-q) \frac{\hat{\beta}^{2}}{2} 
\frac{\partial^{2}}{\partial \hat {\beta}^{2}} + (1-q)^{2} 
\bigg(\frac{\hat{\beta}^{3}}{3}\frac{\partial^{3}}
{\partial \hat {\beta}^{3}} + \frac{\hat{\beta}^{4}}{8} 
\frac{\partial^{4}}{\partial\hat {\beta}^{4}}\bigg)}  
\sum_{l_{i}=0,1} {\cal{D}} \exp(-{\hat{\beta}} {\cal{L}})
\label{rppsd}
\eeq
we obtain the series expansion
\beq
Z_{q}^{(2)}(\beta) = [f(\beta)]^{N} [1-(1-q) {\cal{F}}_{1} + (1-q)^{2} 
                      (
                     {\cal{F}}_{2} + {\cal{F}}_{3} + {\cal{F}}_{4} +
                     {\cal{F}}_{5} )  ],
\label{rplts}
\eeq
where the coefficients read
\bea
{\cal{F}}_{1} &=& 6N  {\hat{\beta}}^{2} \  {\cal{E}}(\beta) +
                    18N(N-1) {\hat{\beta}}^{2} \  [{\cal{E}}(\beta)]^{2},\nn \\
{\cal{F}}_{2} &=& ( 6N {\hat{\beta}}^{4} - 8N
		    {\hat{\beta}} ^{3} ) {\cal{E}}(\beta), \nn \\
{\cal{F}}_{3} &=&  (126N(N-1){\hat{\beta}} ^{4} - 72N(N-1)
		   {\hat{\beta}} ^{3} ) [{\cal{E}}(\beta)]^{2}, \nn \\
{\cal{F}}_{4} &=&  (324N(N-1)(N-2){\hat{\beta}} ^{4}
		     - 72N(N-1)(N-2){\hat{\beta}} ^{3}
		     ) [{\cal{E}}(\beta) ]^{3}, \nn \\
{\cal{F}}_{5} &=& 162N(N-1)(N-2)(N-3) {\hat{\beta}} ^{4}
                   [{\cal{E}}(\beta) ]^{4}, \nn\\
{\cal{E}}(\beta) &=& \exp (- 2 {\hat{\beta}}) [f(\beta)]^{-1}, \qquad
 f(\beta) = 1+3 \exp ( -2 {\hat{\beta}}). 
\label{efs}
\eea
The defining property of the internal energy in the second constraint
framework
\beq
U_{q}^{(2)}(\beta) = - \frac{\partial}{\partial\,\beta}\;
\ln_{q}\,Z_{q}^{(2)}(\beta)   
\label{UDEF}
\eeq
and the series (\ref{rplts}) allows us to extract the perturbative 
result 
\beq
U_{q}^{(2)}(\beta) = {\cal{G}}_{1}+ (1-q) \ ({\cal{G}}_{2}+{\cal{G}}_{3}+
                     {\cal{G}}_{4}),
\label{IELT}
\eeq
where the coefficients may be listed as
\bea
& & {\cal{G}}_{1} = \frac{6 \ N}{\Omega_{R}} \  {\cal{E}}(\beta),  \qquad
 {\cal{G}}_{2} = \bigg(\frac{6N^{2}}{\Omega_{R}} \ln{f(\beta)} +
                    \frac{12N \beta}{\Omega_{R}^{2}}- \frac{12N \beta^{2}}
                    {\Omega_{R}^{3}}\bigg){\cal{E}}(\beta),   \nn \\
& & {\cal{G}}_{3} =  \bigg(\frac{36 N \beta^{2}}{\Omega_{R}} + \frac{36N(N-1) 
                    \beta}{\Omega_{R}^{2}}-\frac{72N(N-1)\beta^{2}}
                    {\Omega_{R}^{3}} \bigg)[{\cal{E}}(\beta)]^{2}, \nn \\
& & {\cal{G}}_{4} = \frac{216N(N-1)\beta^{2}}{\Omega_{R}^{3}}
                    [{\cal{E}}(\beta)]^{3}. \nn
\eea
The thermodynamic quantities in the third constraint scenario 
materialize on transforming $\beta \rightarrow \beta^{\prime}$ 
\textit{\`{a} la} (\ref{trans}). The key quantity $\kappa$ enacting 
this transformation may be factorized as  
\beq
\kappa \equiv \sum_{i} (p_{i}^{(2)})^{q} 
= \chi (\beta) \; (Z_{q}^{(2)}(\beta))^{-q},\qquad
\chi(\beta) =  \sum_{l_{i}=0,1} {\cal{D}} [1-(1-q)
               {\hat{\beta}} {\cal{L}}]^{q/1-q}.
\label{KAP}
\eeq  
In the present case we evaluate the sum $\chi(\beta)$ perturbatively 
and retain terms upto order $(1-q)^{2}$. Recasting as a differential 
series 
\bea
\chi(\beta) &=& 
              \qb{ 1-(1-q)\bigg(\hat{\beta} \frac{\partial}{\partial 
\hat{\beta}} 
           - \frac{\hat{\beta}^{2}}{2} \frac{\partial^{2}}{\partial
	     \hat{\beta}^{2}} \bigg)  \nn \\
            & & + (1-q)^{2}
            \bigg(\hat{\beta}^{2}  \frac{\partial^{2}}{\partial 
\hat{\beta}^{2}} +
            \frac{5 \hat{\beta}^{3}}{6}  \frac{\partial^{3}}{\partial
	    \hat{\beta}^{3}} + \frac{\hat{\beta}^{4}}{8} 
\frac{\partial^{4}}{\partial
	    \hat{\beta}^{4}} \bigg) }  \sum_{l_{i}=0,1} {\cal{D}}
            \exp(- {\hat{\beta}} {\cal{L}} ),
\label{NEPd}
\eea
we obtain $\chi(\beta)$ as
\bea
\chi(\beta) &=& Z_{q}^{(2)}(\beta)+[f(\beta)]^{N} [(1-q) 6N \ {\hat{\beta}}
                {\cal{E}}(\beta)
               +(1-q)^{2} \ 12N \{ ( {\hat{\beta}}^{2}- {\hat{\beta}}^{3})
                {\cal{E}}(\beta)  \nn \\
            & & +3(N-1)\ ( {\hat{\beta}}^{2}- 3{\hat{\beta}}^{3}) 
               [{\cal{E}}(\beta)]^{2} -9(N-1)(N-2)\ {\hat{\beta}}^{3}
               [{\cal{E}}(\beta)]^{3} \}].
\label{NEPs}
\eea
The relations (\ref{KAP}), (\ref{NEPs}) and the partition function 
(\ref{rplts}) now yield $\kappa$ upto the order $(1 - q)^{2}$:
\beq
\kappa = 1 + (1-q) {\cal{P}}_{1} + (1-q)^{2} 
                      ({\cal{P}}_{2}+{\cal{P}}_{3}+{\cal{P}}_{4} ),
\label{EPlt}
\eeq
where the coefficients read
\bea
&  &{\cal{P}}_{1} = N \ln f(\beta) + 6 N {\hat{\beta}} \ 
{\cal{E}}(\beta),\nn \\
&  & {\cal{P}}_{2} = \frac{N^{2}}{2} [\ln f(\beta)]^{2}+
                     (6 N {\hat{\beta}}^{2} - 12 N 
                     {\hat{\beta}}^{3} + 6 N^{2}{\hat{\beta}} 
                     \ln f(\beta) ) \ {\cal{E}}(\beta),  \nn \\
& & {\cal{P}}_{3} = ( 18 N (N-1){\hat{\beta}}^{2} + (108 N - 72 N^{2})
                    {\hat{\beta}}^{3} ) [{\cal{E}}(\beta)]^{2}, \nn \\
& & {\cal{P}}_{4} = 216 N (N-1) {\hat{\beta}}^{3}  
[{\cal{E}}(\beta)]^{3}. 
\label{KAPCO}
\eea
The transformation (\ref{trans}) may now be produced as a perturbative 
series:
\beq
\beta = \beta^{\prime} + (1-q) \ \beta^{\prime}  g(\beta^{\prime}) 
        +(1-q)^{2} \  \beta^{\prime} h(\beta^{\prime}), 
\label{ltrans}
\eeq
where the coefficients read
\bea
g(\beta^{\prime}) &=& N \ln f(\beta^{\prime}) 
+ 12 N {\hat{\beta^{\prime}}} \,{\cal{E}(\beta^{\prime})},\nn\\
h(\beta^{\prime}) &=& 432 N (N-1){\hat{\beta^{\prime}}}^{2}
                      [{\cal{E}(\beta^{\prime})}]^{3}
                       + 18 N {\hat{\beta^{\prime}}}^{2}
                      (5 N - 3 + (12 - 8 N){\hat{\beta^{\prime}}})
                      [{\cal{E}(\beta^{\prime})}]^{2} \nn \\
                  & & + 6N(3{\hat{\beta^{\prime}}}^{2}-4{\hat{\beta^{\prime}}}
                      +N{\hat{\beta^{\prime}}}) \ {\cal{E}(\beta^{\prime})}  
                      +\frac{N^{2}}{2} \ [\ln f(\beta^{\prime})]^{2}.
 \label{TRterm2} 
\eea
To proceed with our computation, we, however, need to invert the series
(\ref{ltrans}). This, upto the order $(1-q)^{2}$ is given as 
\beq
\beta^{\prime}= \beta[1-(1-q)\ g(\beta)+(1-q)^{2} \ ( \beta g(\beta)
                g^{\prime}(\beta) + [g(\beta)]^{2} - h(\beta) ) ]. 
\label{TRlt2}
\eeq
Insofar as the transformation $\beta \rightarrow \beta^{\prime}$ 
a temperature dependent process we can not use the \textit {same} 
construction of the transformation  in the \textit {distinct} regimes 
of high and low temperatures. This was, however, done in the earlier 
work on the rigid rotator [\cite{BST}]. On this we disagree with them.

\par

Employing the key properties (\ref{CZ}) and (\ref{PEQ}) we now extract 
the generalized partition function in the setting of the third 
constraint:
\beq 
[\bar{Z}^{(3)}_{q}(\beta)]^{(1-q)} = \kappa + (1-q)^{2}12 N^{2} 
                                {\hat{\beta}}^{2} 
                                [12 {\hat{\beta}}{\cal{E}}(\beta)+
                                \ln f(\beta) ] \ \{{\cal{E}}(\beta) - 
                                3 [{\cal{E}}(\beta)]^{2} \}.
\label{GP3lt}
\eeq 
The defining thermodynamic relationships
\bea
\beta \frac{\partial U_{q}}{\partial \beta} &=& \frac{\partial \ln_{q}
                                {\bar{Z}^{(3)}_{q}}}{\partial \beta},\\ 
\label{iegp}
C_{q}^{(3)} &=& -k \beta \frac{\partial \ln_{q} \bar{Z}_{q}^{(3)}}
               {\partial \beta}
\label{sh1}
\eea
now allow us to produce the specific heat at the low 
temperature limit in the context of the third constraint:
\beq
C_{q}^{(3)} = {\cal{C}}_{BG} - (1-q)[ {\cal{C}}_{1} + {\cal{C}}_{2} + 
                   {\cal{C}}_{3} + {\cal{C}}_{4} ],
\label{SH3lt}  
\eeq
where the familiar classical Boltzmann-Gibbs extensive term  
${\cal{C}}_{BG}$ reads
\beq
{\cal{C}}_{BG} = 12 N k {\hat{\beta}}^{2} \ {\cal{E}}(\beta)  
               - 36 N k {\hat{\beta}}^{2} \ [{\cal{E}}(\beta)]^{2}.
 \label{BG terms}  \nn
\eeq
The nonextensive terms in (\ref{SH3lt}) at the order of $(1-q)$ 
are listed below: 
\bea
{\cal{C}}_{1} &=& 12Nk [N ( {\hat{\beta}}^{2} - 2
                  {\hat{\beta}}^{3}) \ln f(\beta) + 
                  {\hat{\beta}}^{2} -4  {\hat{\beta}}^{3}
                  +2 {\hat{\beta}}^{4}]  {\cal{E}}(\beta), \nn \\
{\cal{C}}_{2} &=& 36Nk [ N \ln f(\beta) (6{\hat{\beta}}^{3}-
               {\hat{\beta}}^{2}) + (12+2N){\hat{\beta}}^{3} 
               -(14+8N) {\hat{\beta}}^{4}] [{\cal{E}}(\beta)]^{2},
               \nn\\
{\cal{C}}_{3}  &=& 216Nk [{\hat{\beta}}^{3} + 2{\hat{\beta}}^{3}\ln f(\beta)
                   +(10N-12) {\hat{\beta}}^{4} 
                   -4 {\hat{\beta}}^{3}] [{\cal{E}}(\beta)]^{3},\nn\\
{\cal{C}}_{4}  &=& -3888N(N+1) k \  {\hat{\beta}}^{4} \  
                   [{\cal{E}}(\beta)]^{4}.  
\label{CNE}                  
\eea 
In the classical $q\rightarrow1$ limit the specific heat (\ref{SH3lt}) 
of $N$ rotators reduces, expectedly, to its Boltzmann-Gibbs value. In 
the $N = 1$ rotator case, we, unfortunately, fail to notice this 
occuring in [\cite{BST}].  

\par

At the low temperature limit the leading term in internal 
energy may now be computed via integrations in (\ref{iegp}).
We quote the result:
\beq
U_{q}^{(3)} = \U_{BG}+(1-q)\  (\U_{1}+\U_{2}+\U_{3}+\U_{4}+\U_{5}),
\label{UNE}
\eeq
where the coefficients read
\bea
& & \U_{BG} = 6N \Omega_{R} \exp(-2 \hat{\beta}) \ [1-3\exp(-2 \hat{\beta}) +
            6 \exp(-4 \hat{\beta})], \nn \\
& & \U_{1}=12 N \Omega_{R} \exp(-2 \hat{\beta})\ [\hat{\beta}-\hat{\beta}^{2}],
               \nn \\ 
& &  \U_{2}  =  N \Omega_{R} \exp(-4 \hat{\beta})  \ [1+36(N-2) \hat{\beta}+
               18(8N+1)\hat{\beta}^{2}], \nn \\
& & \U_{3} = 2N \Omega_{R} \exp(-6 \hat{\beta}) \ [(2-N)+(11-20N) 6\hat{\beta}
             -(12+21) 18 \hat{\beta}], \nn \\
& & \U_{4} = \frac{81N \Omega_{R}}{8} \ \exp(-8 \hat{\beta}) \ 
              [(6N +3/2)+(14N + 3/2) \hat{\beta} +
             (36 N +42) \hat{\beta}^{2} ], \nn \\
& & \U_{5} = \frac{2916N \Omega_{R}}{125} \ \exp(-10 \hat{\beta})  \
             [(1+10 \hat{\beta})(9N + 4)-200(N + 1) \hat{\beta}^{2}].
\label{UCO}
\eea

%
%
\setcounter{equation}{0}
\section{Gas molecules with translational and rotational 
 degrees of freedom}
\label{DM}
Physically another aspect of nonextensivity may be of interest. Most 
often systems incorporate multiple degrees of freedom. In the 
nonextensive scenario this may indicate that the thermodynamic 
quantities such as specific heat are likely to have temperature dependent 
terms of mixed origin.
This terms are generated due to the \textit {coexistence} of 
different degrees of freedom. Another way of visualizing this is that 
nonextensivity of statistics induces effective interaction terms 
between two separate component of the microscopic Hamiltonian of the 
system. In contrast to this, different degrees of freedom remain 
disjoint 
in the usual extensive statistical mechanics. As a prototype of this 
idea we consider here an ideal diatomic gas endowed with both the 
translational and the rotational degree of freedom. For the 
purpose of simplification we assume the molecules to be of large mass 
so that we may  disregard the vibrational modes. We also restrict
our considerations here to the high temperature limit.   

\par

The energy of a diatomic molecule is given by
\beq
\epsilon = \sum_{i = 1}^{3} \frac{p_{i}^{2}}{2m} + \frac{\hbar^{2}}{2I} 
l(l+1), \qquad l =0,1, \hdots.
\label{eeg} 
\eeq
The generalized partition function in the third constraint is given by
\beq
\bar{Z}^{(3)}_{q} =  \frac{1}{N! \ h^{3N}} \sum_{l_{i}=0}^{\infty} {\cal{D}}
                    \int_{-\infty}^{\infty} \qb{1-(1-q) \frac{\beta}{c}
           \bigg(\frac{P^{2}}{2m}+\frac{{\cal{L}}}{\Omega_{R}}-U_{q}\bigg)}
         ^{\frac{1}{1-q}} \prod_{i=1}^{N} {\rm{d}}^{3}q_{i} 
{\rm{d}}^{3}p_{i},
\label{gpfdg}
\eeq
where $P^{2} = \sum_{i} p_{i}^{2}$. Following Sec. \ref{HTL}, here also
we, in the high temperature limit, replace the summation on the rotational 
levels by an integral: 
\beq
\bar{Z}^{(3)}_{q} = {\cal{V}}\int_{l_{i}=0}^{\infty} \int_{p_{i}=-\infty}
                    ^{\infty} {\cal{D}}\qb{1-(1-q) \frac{\beta}{c}
               \bigg(\frac{P^{2}}{2m}+\frac{{\cal{L}}}{\Omega_{R}}-U_{q}\bigg)}
             ^{\frac{1}{1-q}} \prod_{i=1}^{N} {\rm{d}}l_{i} {\rm{d}}^{3}p_{i},
\label{gpfdgi}
\eeq
where the numerical factor reads ${\cal{V}} = V^{N}/N!\; h^{3 N}$. The 
partition function may now be exactly evaluated as 
\beq
\bar{Z}^{(3)}_{q} =  {\cal{V}}\; c^{5 N/2} (1-q)^{-5 N/2} \qb{\exp_{q}
\bigg(\frac{\beta}{c} U_{q}^{(3)}\bigg)}^{\Lambda_{(5/2)}} \; 
{\sf{G}}\;\; \hat{\beta}^{-N} \check{\beta}^{-3N/2},
\label{gpfdgs}
\eeq
where $\sf{G}$ is a ratio of two gamma functions and 
$\check{\beta}$ is a scaled temperature:
\beq
{\sf{G}} =  {\Gamma \bigg(\frac{2-q}{1-q}\bigg)} 
                \bigg (\Gamma\bigg(\frac{2-q}{1-q}
                 +\frac{5 N}{2}\bigg)\bigg )^{-1}, \qquad
\check{\beta} = \frac{\beta}{2 \pi m}.
\label{g1st}
\eeq
A similar evaluation of the internal energy leads us to 
\beq
U_{q}^{(3)}(\beta) = \frac{5 N}{2}\;\frac{{\cal{V}} \; 
c^{\frac{5N}{2}+1}}
{\beta \bar{Z}^{(3)}_{q}} \; (1-q)^{-5 N/2} \qb{\exp_{q}
\bigg(\frac{\beta}{c} U_{q}^{(3)}\bigg)} ^{\Lambda_{(5/2)}}  
                {\sf{G}} \;\;\hat{\beta}^{-N} 
\check{\beta}^{-3 N/2}.
\label{ies1}
\eeq
Fortunately the ratio of two implicit equations (\ref{gpfdgs}) and 
(\ref{ies1}) assumes a simple form 
\beq
U_{q}^{(3)} = \frac{5 N c}{2 \beta}.
\label{ZUR}
\eeq  
On substituting (\ref{ZUR}) in (\ref{gpfdgs}) and (\ref{ies1}) we 
evaluate the generalized partition function and the internal 
energy explicitly: 
\bea
\bar{Z}^{(3)}_{q} &=& \Psi \;[\exp_{q}(5 N/2)]^{\mu_{5/2}} 
                    \hat{\beta}^{-N \mathfrak{L}_{(5/2)}}
              \check{\beta}^{-\frac{3N}{2} \mathfrak{L}_{(5/2)}},\nn\\
U_{q}^{(3)}(\beta) &=& \frac{5N}{2 \beta} \;  \Psi^{(1-q)} \;
                     \Lambda_{(5/2)}^{\mu_{5/2}} \;
                    \hat{\beta}^{(q-1)N \mathfrak{L}_{(5/2)}} \;
          \check{\beta}^{(q-1)\frac{3 N}{2} \mathfrak{L}_{(5/2)}},\nn\\
\Psi &=& [{\cal{V}} \; (1-q)^{-\frac{5 N}{2}} \;  
       {\sf{G}}]^{\mathfrak{L}_{(1)}}.
\label{ieef1}
\eea
The specific heat of the diatomic gas in the high temperature limit is 
directly produced via the internal energy in (\ref{ieef1})
\beq
 C_{q}^{(3)}(\beta) \equiv \frac{\partial}{\partial\; T}\;  
U_{q}^{(3)}(\beta) 
= \frac{5 N k }{2} \; {\mathfrak{L}}_{(5/2)} \;
                     \Lambda_{(5 N/2)} ^{\mu_{5/2}} \;
                     \Psi^{{\mathfrak{L}}_{(5/2)}} \;
         {\check{\beta}}^{(q-1)\frac{3N}{2} {\mathfrak{L}}_{(5/2)}} \;
                {\hat{\beta}}^{(q-1)N {\mathfrak{L}}_{(5/2)}}.
\label{shdg}
\eeq  
The above specific heat of the diatomic gas can also be verified by 
employing the $\beta \rightarrow \beta^{\prime}$ transformation 
procedure discussed and used in Sec. \ref{LTL}. The 
relevant transform in the present case is quoted below:
\beq
\frac{1}{\beta^{\prime}} = \qb{{\cal{V}} \; \Omega_{T}^{3 N/2}
\Omega_{R}^{N} \; {\sf{G}}\; 
(1-q)^{-5N/2}}^{(1-q){\mathfrak{L}}_{(5/2)}} \;
                           \Lambda^{2{\mathfrak{L}}_{(5/2)}} \;
                       \qb{\frac{1}{\beta}}^{{\mathfrak{L}_{(5/2)}}}. 
\label{tedg}
\eeq
Expanding the specific heat (\ref{shdg}) of the gas molecules  
having both translational and rotational degrees 
of freedom in the parameter $(1-q)$, we retain terms upto the order of 
$(1-q)^{2}$:
\beq 
C_{d} = \frac{5 N }{2} k  \qb{1 + (1-q) \Xi_{d} + (1-q)^{2} 
      \bigg(\frac{\Xi_{d}^{2}}{2} + \frac{5 N}{2} \Xi_{d} 
      -\frac{25 N^{2}}{8} - \frac{5 N}{4} \bigg) }, 
\label{SHTSg}
\eeq
where $\Xi_{d} = \ln {\cal{V}} - (3 N/2) \ln {\check{\beta}} 
             - N \ln{\hat{\beta}} + 5 N$.
For the purpose of comparison we now consider an expansion of the 
translational specific heat [\cite{SA1}] in the parameter $(1 - q)$
while retaining terms upto order $(1-q)^{2}$: 
\beq
C_{t} =  \frac{3N}{2} k  \qb{1 + (1-q) \Xi_{t} + (1-q)^{2} 
      \bigg(\frac{\Xi_{t}^{2}}{2} + \frac{3N}{2} \Xi_{t} 
      -\frac{9N^{2}}{8} - \frac{3N}{4} \bigg) },
\label{SHTSt}
\eeq
where $\Xi_{t} = \ln {\cal{V}} - (3 N/2)\; \ln {\check{\beta}} + 3 N.$
Applying the same expansion in the rotational specific heat in the high 
temperature case (\ref{shht}), we obtain 
\beq
C_{r} =  N \;  k  \qb{1 + (1-q) \Xi_{r} + (1-q)^{2} 
      \bigg(\frac{\Xi_{r}^{2}}{2} + N \  \Xi_{r} 
      -\frac{N^{3}}{2} - \frac{N^{2}}{2} \bigg)},
\label{SHTSr}
\eeq
where $\Xi_{r} = -N \ln{\check{\beta}} + 2 N$. Summarizing the above 
expansion in the region $q \rightarrow 1$ we observe that 
the specific heat of the gas molecules with both translational and 
rotational degree of freedom may be written as the following sum:
\beq
C_{d} = C_{t} + C_{r} + M_{t} + M_{r} + M_{t,r},
\label{Mixing}
\eeq
where the additional terms on the rhs read
\bea
M_{t} &=& (1-q) Nk \ln\bigg( \frac{\cal{V}}{{\check{\beta}}^{3N/2}}\bigg)
          +(1-q)^{2} \frac{Nk}{2} \ln\bigg( \frac{\cal{V}}{{\check{\beta}}
          ^{3N/2}}\bigg) \bigg[1+ 24N \ln\bigg( \frac{\cal{V}}
          {{\check{\beta}}^{3N/2}}\bigg) \bigg],
   \label{Mixingt}  \nn\\
M_{r} &=& -(1-q) \frac{3N^{3} k}{2} \ln{\hat{\beta}} + (1-q)^{2} 
          \frac{N^{3}k}{4} \bigg[ 3 (\ln{\hat{\beta}})^{2} - 53 
          \ln{\hat{\beta}}\bigg],
   \label{Mixingr}\\
M_{t,r} &=& (1-q)6N^{2}k + (1-q)^{2} N^{2}k  \bigg[\frac{315N}{8} - 2 - N^{2}
            -\frac{15N}{4}\ln{\check{\beta}} \ln{\hat{\beta}} -\frac{5}{2}
            \ln{\cal{V}} \ln{\hat{\beta}} \bigg].
   \label{Mixingtr}  \nn
\eea
Evidently the specific heat $C_{d}$ of the diatomic molecules is 
\textit {not} a linear sum of the translational specific heat $C_{t}$ 
and the rotational specific heat $C_{r}$. The additional terms 
owing their origin to nonextensivity may be divided into three classes.
The terms $M_{t}$ and $M_{r}$ in (\ref{Mixing}) contain the effects of 
purely translational and purely rotational degrees of freedom, 
respectively. Qualitatively, these terms may be described as a kind
of \textit {nonlinear} effect of mixing of degrees of freedom. On the
other hand, the term $M_{t,r}$ incorporate the effect of \textit {both} 
the degrees of freedom. These cross terms indicate the existence of some 
sort of effective interaction between individual degrees of freedom, 
such as the translational and the rotational examples discussed here.
These effective interactions are temperature dependent and may be 
thought of as induced by the nonextensive statistics. This 
qualitatively distinct feature is absent in Boltzmann-Gibbs statistics.   
       
%
%
\setcounter{equation}{0}
\section{Concluding Remarks}
\label{CR}
We have studied the specific heat of $N$ rigid rotators in the high 
temperature limit. Using the disentanglement of the $q$ exponential in 
an infinite product series of ordinary exponentials, we have developed a 
pertubative method to calculate the specific heat upto any particular 
order in the nonextensivity parameter $(1-q)$. This procedure is applied 
to find the specific heat of $N$ rotators in the low temperature limit. 
Finally we calculate the specific heat of a diatomic gas with a 
combination of the translational and the rotational degrees of freedom.  
The $q \rightarrow 1$ limiting value of the specific heat specific heat
leads us to conclude that there is a mixing between the different kinds of 
degrees of freedom due to nonextensive statistics. This mixing of 
independent degrees of freedom may be visualized as an effective 
interaction between them owing its origin due to nonextensivity of 
statistics. A study of these problems in the context of two parametric  
entropies [\cite{AMSTW},\cite{VSCT}] should be worth pursuing.

%
%
%
\section*{Acknowledgements} 
R. Chandrashekar and S.S. Naina Mohammed would like to acknowledge 
the fellowships received from the Council of Scientific and Industrial 
Research (India), and the University Grants Commission (India), 
respectively. These authors would like to thank A.M. Mathai for 
inviting us to visit CMS Pala Campus (India) where part of the work 
was done. 
%

%
%
%
      

\begin{thebibliography}{99}
\bibitem{CT1} C. Tsallis, J. Stat. Phys {\bf 52}, 479 (1988).  
%
\bibitem{PP} A.R. Plastino and A. Plastino, Phys. Lett. {\bf A174}, 384 
(1993).
%
\bibitem{FL} K.S. Fa and E.K. Lenzi, J. Math. Phys {\bf 42}, 1148 (2001).
%
\bibitem{DJ} D. Jiulin, Phys. Lett. {\bf A320}, 347 (2004).
%
\bibitem{KLQ1} G. Kaniadakis, A. Lavagno and P. Quarati, Phys. Lett. {\bf B369}
308 (1996).
%
\bibitem{KLQ2} G. Kaniadakis, A. Lavagno and P.Quarati, Astrophys. Space Sci.
{\bf 228}, 77 (1995).
%
\bibitem{TBM} C. Tsallis, G. Bemski and R.S. Mendes, Phys. Lett. {\bf A257},
93 (1999).
%
\bibitem{AH} E. Akturk and A. Harkin, {\it Nonextensive statistical mechanics 
application to vibrational dynamics of protein folding}, cond-mat/0703408.
%
\bibitem{CUR} E.M.F. Curado and C. Tsallis, J. Phys. {\bf A24}, L69 (1991).
%
\bibitem{TMP} C. Tsallis, R.S. Mendes and A.R. Plastino, Physica {\bf A261}, 
534 (1998).
%
\bibitem{SA1} S. Abe, Phys. Lett. {\bf A263}, 424 (1999).
%
\bibitem{AMPP} S. Abe, S. Martinez, F. Pennini and A. Plastino, Phys. Lett. 
{\bf A281}, 126 (2001).
%
\bibitem{BCCT}  B.J. Cabral and C. Tsallis, Phys. Rev. {\bf E66},
065101 (2002).
%
\bibitem{BST} G.B. Bagci, R. Sever and C. Tezcan, Mod. Phys. Lett 
{\bf B18} No. 11, 467 (2004).
%
\bibitem{CQ} C. Quesne, Int. J. Theor. Phys. {\bf 43}, 545 (2004).
%
\bibitem{AMSTW} A.M. Scarfone and T. Wada, Phys. Rev. {\bf E72},
026123 (2005).
%
\bibitem{VSCT} V. Schwammle and C. Tsallis, {\it Two parameter generalization 
of the logarithm and the exponential functions and Boltzmann-Gibbs-Shannon 
entropy}, cond-mat/0703792.
\end{thebibliography}
\end{document}